\documentclass{jfm}

\usepackage{graphicx}
\usepackage{natbib}

\usepackage{color}
\usepackage{soul}
\usepackage{booktabs}
\ifCUPmtlplainloaded \else
  \checkfont{eurm10}
  \iffontfound
    \IfFileExists{upmath.sty}
      {\typeout{^^JFound AMS Euler Roman fonts on the system,
                   using the 'upmath' package.^^J}%
       \usepackage{upmath}}
      {\typeout{^^JFound AMS Euler Roman fonts on the system, but you
                   dont seem to have the}%
       \typeout{'upmath' package installed. JFM.cls can take advantage
                 of these fonts,^^Jif you use 'upmath' package.^^J}%
      }
  \else
  \fi
\fi

\ifCUPmtlplainloaded \else
  \checkfont{msam10}
  \iffontfound
    \IfFileExists{amssymb.sty}
      {\typeout{^^JFound AMS Symbol fonts on the system, using the
                'amssymb' package.^^J}%
       \usepackage{amssymb}%

      }{}
  \fi
\fi

\ifCUPmtlplainloaded \else
  \IfFileExists{amsbsy.sty}
    {\typeout{^^JFound the 'amsbsy' package on the system, using it.^^J}%
     \usepackage{amsbsy}}
    {}
\fi

\newsavebox{\astrutbox}
\sbox{\astrutbox}{\rule[-5pt]{0pt}{20pt}}

\newlength{\halfwidth}
\title[A note on acoustic turbulence]%
{ A note on acoustic turbulence}

\author[ E. Lindborg]%
{ Erik Lindborg  \ns }

\affiliation{ KTH Mechanics,
SE-100 44 Stockholm, Sweden}

\pubyear{2013}
\volume{???}
\pagerange{???--???}
\date{?; revised ?; accepted ?. - To be entered by editorial office}
\begin{document}

\maketitle

\begin{abstract} 
We consider a three-dimensional acoustic field of an  ideal gas in which all entropy production is confined to weak shocks and  show that similar scaling relations hold for such a field as for forced Burgers turbulence, where 
the shock amplitude scales as 
 $  (\epsilon   d)^{1/3} $ and the $ p $:th order structure function scales as $ (\epsilon  d)^{p/3} r/d$,  $ \epsilon $ being the  mean energy dissipation per unit mass, $ d $ the mean  distance between the shocks and  $ r $ the separation distance.
However,  for the acoustic field $ \epsilon $ should be replaced by  $  \epsilon + \chi $, where $ \chi $ is  associated with entropy production due to heat conduction.  In particular, the third order longitudinal structure function scales as
$ \langle \delta u_r^3 \rangle = -C(\epsilon + \chi) r $, where $ C $ takes the value $ 12/5(\gamma +1) $  in the weak shock limit,  $ \gamma = c_p/c_v$ being  the ratio between the specific heats at constant pressure and constant volume.
\end{abstract}

\section{Introduction}
A random acoustic field in which many modes are excited may be characterised as acoustic turbulence. A long standing problem in turbulence theory is to determine the statistical properties of such a field. Two approaches have been developed to tackle this problem. 
 \cite{ZakharovSagdeev1970} used weak turbulence theory to derive an energy wave number spectrum of the form $ \sim k^{-3/2} $ in three dimensions.  This approach was criticised by  \cite{KadomtsevPet1973}, who argued that shocks will always develop in an acoustic field
if the Reynolds number is sufficiently large, and give rise to an energy spectrum of the form $ \sim k^{-2} $. 
Insightful discussions on the arguments in favour of each of the two approaches are given by \cite{FalkovichMeyer1996} and \cite{Lvov}. In this note, we will not add to these discussions.
Instead, we will assume that there exist a regime in which nonlinearities are sufficiently strong for shocks to form in such an abundance that virtually all entropy production is confined to  shocks, and yet sufficiently weak for the shocks to be weak.  The aim of the paper is to derive  scaling relations for such a hypothetical regime, which may be tested experimentally or numerically.  Before doing this, it may be illustrative to consider forced Burgers turbulence as a simple  example of shock dominated turbulence.  

We consider statistically stationary solutions to Burgers equation \cite[]{Burgers1948},
\begin{equation} \label{Burger}\frac{\partial u}{\partial t} + u \frac{\partial u}{\partial x} = \nu \frac{\partial ^2 u}{\partial x^2} + f \, ,
\end{equation}
in a domain with length $ L $, where $ f $ is a random forcing  with characteristic length scale $ L_f $ and $ \nu $ is the kinematic viscosity.  
Building on previous investigations it can be shown that apart from $ \nu $ the two other parameters characterising forced Burgers turbulence are the mean energy dissipation rate per unit mass, $ \epsilon $, and the mean distance between the shocks, $ d = L/n $, where $ n $ is the number of shocks.
 We define the $ p $:th order velocity structure function as $ \langle | \delta u |^p \rangle $, where $ \delta u = u(x+r)-u(x) $ and $ \langle \rangle $ is the domain average. If $ p $ is an odd integer we can also define a structure functions without taking the absolute value of $ \delta u $, for example the third order structure function $ \langle \delta u^3 \rangle $.  As shown by \cite{BouchaudMezardParisi1995},  \cite{Weinanetal1997} and \cite{Weinan} structure functions of order $ p >  1 $ can be calculated by only taking those increments into account for which the  two points $ x $ and $ x+r $ lie on different sides of a shock. If $ \delta x \ll r \ll d $, where $ \delta x $ is the shock width, the probability that a shock is crossed for an arbitrary increment is equal to $ r n/L = r/d $.  The $ p $:th order structure function can thus be calculated as
\begin{equation} \label{Pth}
\langle | \delta u | ^p \rangle = \langle \Delta u^p \rangle_s \frac{r}{d} \, ,
\end{equation} 
where $ \Delta u $ is the shock amplitude, taken to be positive by definition,  and 
\begin{equation}
 \langle  \Delta u^p \rangle_s  = \frac{1}{n} \sum_{i=1}^{n}  \Delta u _i  ^{p} \, ,
\end{equation} 
 is the average over all shocks.  The expression $ \Delta u^p $ should be read as $ (\Delta u)^{p} $. For simplicity, we omit the parantheses. 
The velocity increment $ \delta u $ is always negative over a shock. The third order structure function can thus be written as 
\begin{equation} \label{Third}
\langle \delta  u^3 \rangle  = - \langle \Delta u ^3 \rangle_s \frac{r}{d} \, .
\end{equation} 
It is straightforward to derive the 
Burgers equation analogue of `the four-fifths law'  of 
\cite{Kolmogorov1941}. Under the assumption of statistical homogeneity and stationarity the following equation is easily derived,
\begin{equation} \label{K1}
\frac{\partial}{\partial r} \langle \delta u^3 \rangle = 
 -12 \epsilon + 6 \nu \frac{\partial^2}{\partial r^2} \langle \delta u^2 \rangle + 6 \langle \delta u \delta f \rangle \, .
\end{equation} 
After integration of (\ref{K1}) the two last terms can be neglected if $ \delta x \ll r \ll L_f $ and we obtain 
\begin{equation} \label{Kolm2} 
  \langle \delta u^3 \rangle = -12 \epsilon  r ,
\end{equation} 
 which was given by \cite{Weinan} and \cite{FalkoSreen}.
Combining (\ref{Third}) and (\ref{Kolm2}) we find
\begin{equation} \label{Amplitude}
\langle \Delta u^3 \rangle_s = 12 \epsilon d \, ,
\end{equation} 
a relation which was derived by \cite{Weinan}.
The $ p $:th order structure function can now be written as
\begin{equation} \label{Struc}
\langle | \delta u | ^p \rangle  = K_p \epsilon^{p/3}d^{p/3-1} r ,
\end{equation} 
where 
\begin{equation}
K_p =   12^{p/3} \frac{ \langle \Delta u^p \rangle_s}{\langle \Delta u^3 \rangle_s^{p/3}} \, ,
\end{equation} 
are non-dimensional prefactors that can be calculated from the probability density function of the shock amplitude $ \Delta u $, if that is known. The kinetic energy spectrum can be calculated as the  Fourier transform of $ - \langle \delta u^2 \rangle/4 $. Using the theory of generalised Fourier transforms \cite[]{Lighthill} we obtain
\begin{equation} \label{Spectrum}
E(k) = \frac{K_2}{2\pi} \epsilon^{2/3} d^{-1/3} k^{-2} \, .
\end{equation} 
The $ k^{-2} $-spectrum was derived by \cite{Burgers1948} in the one-dimensional case and by \cite{Kuznetsov} in the three-dimensional case, without including the parametric dependence on 
 $ \epsilon $ and $ d $.  An interesting property  of  the expressions (\ref{Struc}) and (\ref{Spectrum}) is that they are invariant under superposition of two fields, which follows from the fact that both (\ref{Pth}) and (\ref{Third}) fulfil this type of invariance. 
As discussed in \cite{Frisch}, Landau \cite[]{LandauLif} made the objection  against  \cite{Kolmogorov1941a} that structure functions cannot be universally dependent on $ \epsilon $ and $ r $, since such a law would not be invariant under superposition of two fields. This objection cannot be raised against (\ref{Struc}) and (\ref{Spectrum}), since the prefactors $ K_p $ will adjust in such a way that  this invariance is fulfilled.

Recently, \cite{AugierMohananLindborg} showed that that similar scaling relations are valid for two-dimensional shallow water wave turbulence as for forced Burgers turbulence.  For shallow water wave turbulence  a distinction should be made between structure functions involving increments of the longitudinal and transverse velocity components, just as in the case of two- or three-dimensional incompressible turbulence. \cite{AugierMohananLindborg} showed that the ratio between the longitudinal and transverse structure function of a particular order can be determined from the condition that the velocity step at a shock is confined to the shock normal component. It was also shown that the shock width scales as 
\begin{equation} \label{Width} 
\delta x \sim \frac{\nu} {\; (\epsilon d)^{1/3}} \, .
\end{equation} 
It can be argued that (\ref{Width}) also holds for Burgers turbulence. We will now consider the case of three-dimensional shock dominated acoustic turbulence

\section{Scaling relations for shock dominated acoustic turbulence}

We consider an ideal three-dimensional homogeneous isotropic acoustic field  in a domain with volume $ \cal V $. We assume that the velocity field is irrotational and that all entropy production in the field is confined to weak shocks which we consider as smooth surfaces whose  total area we denote by $ A $.   Since it is assumed that the shocks are weak we can also assume that they interact weakly \cite[]{Apazidis} and cross each other without strong reflections, just as weak shallow water wave shocks \cite[]{AugierMohananLindborg}. Their radius of curvature will therefore be larger than the characteristic distance between them. We define the linear mean distance between the shocks as  $ d \equiv L/n $, where $ L $ is the length of a  straight line segment passing through the domain and $ n $ is the number of shocks that the segment crosses. By the assumptions of isotropy and homogeneity all segments of all straight lines will give the same value of $ d $, provided that sufficiently many chocks are crossed.  A structure function of a scalar flow variable can be calculated just as in the one-dimensional case, as the average over a line segment of a moment of the increment,  for example the density increment $ \delta \rho = \rho({\bf x} + {\bf r} ) - \rho({\bf x}) $, where $ {\bf r} $ is the separation vector. The density structure function of order $ p $ can thus be calculated as
\begin{equation} \label{S1}
\langle | \delta \rho | ^{p}  \rangle_l =   \langle \Delta \rho ^p \rangle_{s}  \frac{r}{d} \, ,
\end{equation} 
where $ \langle \rangle_l $ is the line average, $\Delta \rho $ is the density step at a shock, which we take to be positive by definition,  and $ \langle \rangle_s $ is the average over all shocks that are crossed by the segment.
 
A structure  function  can also be calculated as a domain average which we denote by $ \langle \rangle $, without any subscript. By only taking those increments into account for which $ {\bf r} $ crosses a shock, we find
\begin{equation}
\langle | \delta \rho | ^{p}  \rangle   = \frac{1}{{\cal{V}}}  \int_{\cal{V}} | \delta \rho |^{p}\,  {\mbox{d}} V =  \frac{1}{{\cal{V}}}  \int_{A}  \Delta \rho ^{p} r \cos \theta \, {\mbox{d}} A \, ,
\end{equation} 
where $ \theta $ is the angle between $ {\bf r} $ and the shock normal unit vector, $ {\bf n} $,  defined in such a way that $ \theta \in [-\pi/2, \pi/2] $.  By the assumption of isotropy $ \theta $ can be regarded as a random variable with 
probability density $ |\sin \theta |/2 $.  We thus find
\begin{equation} \label{S2}
\langle | \delta \rho | ^{p}  \rangle    = \langle \cos \theta \rangle_\theta \langle \Delta \rho  ^{p} \rangle_{\!A} \frac{r A }{{\cal{V}}} = \frac{1}{2}  \langle \Delta \rho  ^{p} \rangle_{\!A}    \frac{r A }{{\cal{V}}} ,
\end{equation} 
where 
\begin{equation}
 \langle  \Delta \rho ^{p} \rangle_{\!A} = \frac{1}{A} \int_A \Delta \rho^p \,  {\mbox{d}} A \, , 
\end{equation} 
 is the average over the total shock area and 
\begin{equation}
\langle f(\theta) \rangle_{\theta}  =  \frac{1}{2} \int_{-\pi/2}^{\pi/2} | \sin \theta | f(\theta) \, {\mbox{d}} \theta \, .
\end{equation} 
Evidently,   the domain average   (\ref{S2}) must be equal to the   line average (\ref{S1})  and $ \langle  \Delta \rho ^p \rangle_{\!A} $ must be equal to  $ \langle \Delta \rho  ^{p} \rangle_{s} $.   Therefore,  we must also have \footnote{The corresponding relation 
for an isotropic field in two dimensions is $ {\cal A}/{\cal{L}} = 2d/\pi $, where
$ \cal A $ is the domain area and $ {\cal{L }}$ is the total shock length. For a non-isotropic field the mean crossing distance, $ d $, will be different for different lines. A regular grid of squares with side $ a  $  has $ {\cal A}/{\cal{L}} = a/2 $. A line which is aligned with the grid  has $ d = a $, while a line at fortyfive degrees angle to the grid  has $ d = a/ \sqrt{2} $. }
\begin{equation} \label{Magic}
\frac{{\cal V}}{A} = \frac{1}{2}  d\, ,
\end{equation} 
a relation which will be used in the following.

The velocity step at a shock, $ \Delta u $, which we take to be positive by definition, is confined to the shock normal component.  A shock crossing velocity increment can thus be resolved as 
$ \delta {\bf u} =  \delta u_r  {\bf e}_r +  \delta u_{\theta} {\bf e}_{\theta} $, 
where $ \delta u_r = - \cos\theta \Delta u $, $ \delta u_\theta = \sin \theta \Delta u $, $ {\bf e}_r = {\bf r}/ r $  and $ {\bf e}_{\theta} $ is orthogonal to both $ {\bf e}_r $ and $ {\bf n} \times {\bf e}_{r}  $. 
The structure function $ S_{mn} = \langle | \delta u_{r} | ^m  | \delta u_{\theta} | ^n \rangle $ can be calculated as
\begin{equation}
S_{mn} = 2 \langle |\cos \theta | ^{m+1} |\sin  \theta |^ n \rangle_{\theta} \langle   \Delta u^{m+n} \rangle_{\!A} \frac{r}{d} \, .
\end{equation} 
If $ m $ and $ n $ are integers we can also calculate the structure function $   \langle  \delta u_r   ^{m}   \delta u_\theta^{n}   \rangle  $, which is equal to zero if $ n $ is odd, equal to $ S_{mn} $ if both $ m $ and $ n $ are even and equal to $ - S_{mn} $ if $ m $ is odd and $ n $ is even, since $ \delta u_r $ is always negative over a shock  \cite[]{AugierMohananLindborg}. The second and third order structure functions are of particular interest, 
\begin{eqnarray} \label{SecondS} 
\langle \delta u_r^2 \rangle =  \langle \delta u_\theta ^2 \rangle = \frac{1}{2}  \langle \Delta u^2 \rangle_{\!A} \frac{r}{d} \, ,\\ \label{ThirdS} 
\langle \delta u_r^3 \rangle = \frac{3}{2} \langle \delta u_r \delta u_\theta^2 \rangle = - \frac{2}{5} \langle \Delta u^3 \rangle_{\!A} \frac{r}{d} \, .
\end{eqnarray} 
The radial increment,  $ \delta u_{r} $, is, of course, nothing else than the longitudinal increment, $ \delta u_L $, of standard incompressible turbulence theory \cite[]{Frisch}. However, the increment $ \delta u_{\theta} $ should not be confused with an arbitrary transverse increment $ \delta u_T $, since the subscript $ \theta $ indicates a specific transverse direction in the vicinity of a shock. To obtain $ \langle \delta u_L^{m} \delta u_{T}^n  \rangle $  one should multiply $ \langle \delta u_r^{m} \delta u_{\theta}^n  \rangle $ by $ \int_{0}^{2\pi} \sin^{n} \!\!\phi \, {\mbox{d}} \phi /2 \pi $.
For example, $  \langle \delta u_{T}^2 \rangle  =  \langle \delta u_{\theta}^{2} \rangle/2 $, and the first equality in
 (\ref{SecondS}) is therefore consistent with 
\begin{equation} \label{Isotropic} 
\langle \delta u_L^2 \rangle = \frac{{\mbox{d}}}{{\mbox{d}} r} ( r \langle \delta u_T^2 \rangle ) \, ,
\end{equation} 
which holds for a three-dimensional irrotational isotropic velocity field (see Appendix A). From (\ref{ThirdS})  we also get $  \langle \delta u_ L \delta u_{T}^2 \rangle =   \langle \delta u_ L ^{3} \rangle/3 $, which is only a kinematic consequence of the shock structure, since -- generally -- there are two independent invariants of the third order tensor structure function of an irrotational isotropic field. This is shown in Appendix A.

Assuming that the shocks for most of their life time are in a quasi stationary state we can
use the shock relations for an ideal gas together with the entropy equation to relate $  \langle \Delta u ^3 \rangle_{\!A} $  to the mean entropy production over the shocks. 
For weak shocks, the jump of a flow quantity can be expanded in terms of the shock strength, defined as 
$ z =  (p_2 - p_1 )/{p_1}  $, 
where $ p_1 $ and $ p_2 $ are the pressures before and and after the shock respectively. The jumps in velocity, density, temperature and specific entropy can be expanded as \cite[page 176]{Whitham70}, 
\begin{eqnarray} \label{DU}
\frac{\Delta u}{c_1} & = & \frac{z}{\gamma} + {\cal{O}}(z^2) \, ,  \\
\label{Drho} \frac{\Delta \rho}{\rho_1} & =  & \frac{z}{\gamma} + {\cal{O}}(z^2)  \, ,
\\ \label{DT}
\frac{\Delta T}{T_1}  & = & \frac{\gamma -1}{\gamma} z  + {\cal{O}}(z^2) \, ,\\ \label{DS}
\frac{\Delta S} {c_v} & =  & \frac{\gamma^2 -1}{12\gamma^2} z^3  + {\cal{O}}(z^4) \, , 
\end{eqnarray} 
where $ c_1 $ is the speed of sound before the shock and $ \gamma = c_{p}/c_v $ is the ratio of the specific heats at constant pressure and constant volume.  Assuming that each shock is in a quasi stationary state, the jump in specific entropy can also be calculated by integrating the entropy equation \cite[page 189]{Whitham70}, 
\begin{equation}  \label{dS} 
 \Delta S = \frac{1}{Q} \int_0^{\delta x} \left \{   \frac{ \kappa c_{p} \rho } {T^2}  \left ( \frac{\partial T} {\partial x_i} \right ) ^2 + \frac{\varepsilon}{  T} \right \} {\mbox{d}} x \, ,
\end{equation}
where  $ Q $ is the mass flux per unit area over the shock  in the frame of reference where the shock is at rest,  $ x $ is the local shock normal coordinate,
 $ \kappa $ is the thermal diffusivity and  $ \varepsilon $ is the 
kinetic energy dissipation rate per unit volume.  Since the Mach number is assumed to be close to unity and density fluctuations are assumed to be small we can make the approximation  $ Q   \approx \rho_0 c_0 $, where 
$ \rho_0 $ and $ c_0 $  are background reference values of the density and the speed of sound, respectively,  which we take as the mean values over the whole field.
 Using $  c^2 = \gamma R T $, where $ R $ is the ideal gas constant,  replacing $ c $ by $ c_0 $, $ \rho $ by $ \rho_0 $   and averaging over the total shock area, we obtain
\begin{equation}
\langle \Delta S \rangle_{\!A} \approx \frac{1}{A} \int_{A}  \int_{0} ^{\delta x}  \frac {R\gamma}{c_0^3} \left \{ \frac{\kappa c _p } { T}  \left ( \frac{\partial T} {\partial x_i} \right ) ^2 + \frac{\varepsilon}{\rho_0} \right \} {\mbox{d}} x   {\mbox{d}} A \, .
\end{equation} 
Assuming that the entropy production is confined to the shocks, the domain of integration can be extended to include the whole field, and we get
\begin{equation} \label{dS2}
\langle \Delta S \rangle_{\!A} \approx  \frac{R \gamma }{c_0^3}  (\epsilon + \chi) \frac{\cal{V}}{A} = \frac{R \gamma }{c_0^3}  (\epsilon + \chi) \frac{d}{2}\, ,
\end{equation} 
where $ \epsilon \equiv \langle \varepsilon / \rho_0 \rangle $ is the mean kinetic energy dissipation rate per unit mass and
\begin{equation}
\chi \equiv   \kappa c_p \left \langle   \frac{1}{T}  \left (\frac{\partial T}{\partial x_i} \right ) ^2   \right \rangle    \, .
\end{equation} 
We can now estimate $ \langle \Delta u^{3} \rangle_{\!A} $ to leading order in $ z $ by taking the cube of (\ref{DU}), replacing $ c_1 $ by $ c_0 $, averaging over all shocks  and using (\ref{DS}) and (\ref{dS2}),
\begin{equation} \label{Amplitude2}
\langle \Delta u ^3 \rangle_{\!A}  \approx \frac{6}{\gamma + 1} (\epsilon + \chi) d \, ,
\end{equation} 
which is analogous to relation (\ref{Amplitude}) derived by \cite{Weinan} for Burgers turbulence. 
Inserting this expression into (\ref{ThirdS}) we obtain
\begin{eqnarray} \label{EL}
\langle \delta u_r^3 \rangle  =  - C (\epsilon + \chi) r \, ,
\end{eqnarray} 
where $ C  $ is a positive constant of the order of unity.  
In the weak shock limit we obtain $ C = 12/5(\gamma + 1) $, giving $ C = 9/10 $ for a monoatomic gas and $ C = 1 $ for a diatomic gas.  Equation (\ref{EL}) is similar to the `four-fifths law',  $ \langle \delta u_r^3  \rangle = -(4/5) \epsilon r $,  \cite[]{Kolmogorov1941} for incompressible turbulence  and the corresponding law (\ref{Kolm2}) for Burgers turbulence \cite[]{Weinan, FalkoSreen}, with the difference that  $ \epsilon $ is replaced by $ \epsilon + \chi $. 

From (\ref{Amplitude2}) we can conclude that the  shock amplitude scales as $ \Delta u \sim (\epsilon + \chi)^{1/3} d^{1/3} $ and that a structure function of order $ p > 1 $, scales as
 $ \langle | \delta u_r |^{p} \rangle \sim (\epsilon + \chi)^{p/3} d^{p/3 - 1} r $.
The  structure functions $ \langle \delta {\bf u} \cdot \delta {\bf u} \rangle = \langle \delta u_{r}^2 \rangle + \langle \delta u_{\theta}^2 \rangle $ and $ \langle (c_0 \delta \rho /\rho_0)^2 \rangle $, are of particular interest, as are 
 $ \langle \delta u_r   \delta {\bf u} \cdot \delta {\bf u} \rangle $ and $ \langle \delta u_r  (c_0 \delta \rho /\rho_0)^2 \rangle $. Replacing the local density $ \rho_1 $ with $ \rho_0 $ in (\ref{Drho}) and the local speed of sound in (\ref{DU}) with $ c_0 $ will only give rice to alterations of $ {\cal{O}} (z^{2}) $ on the right hand sides. To leading order in $ z $ we thus find
\begin{eqnarray} \label{Second} 
 \langle (c_0 \delta \rho /\rho_0)^2 \rangle =  \langle \delta {\bf u} \cdot \delta {\bf u} \rangle \sim (\epsilon + \chi)^{2/3} d^{-1/3} r \, , \\ \label{Fluxes}
 \langle \delta u_r  (c_0 \delta \rho /\rho_0)^2 \rangle =   \langle \delta u_r   \delta {\bf u} \cdot \delta {\bf u} \rangle   = -\frac{5}{3} C (\epsilon + \chi) r \, ,
\end{eqnarray} 
where the expression for $  \langle \delta u_r  (c_0 \delta \rho /\rho_0)^2 \rangle $ is similar to  the \cite{Yaglom} relation for the third order velocity-scalar structure function of incompressible turbulence.
The second order velocity structure function is associated with kinetic energy and the density structure function with the energy form which has been referred to as `acoustic potential energy' \cite[page 13]{LighthillW}. These two forms of energy are equipartitioned in a linear acoustic wave. Quite interestingly, to leading order in $ z $, equipartition also holds for a field of weak shocks, and if the third order structure functions are supposed to be associated with energy fluxes, the relation (\ref{Fluxes}) indicates that kinetic and potential energy fluxes are equipartitioned, just as in shallow water wave turbulence \cite[]{AugierMohananLindborg}.
The kinetic and potential energy spectra scale as
\begin{equation} \label{Spectra} 
E_K(k) = E_P (k ) \sim (\epsilon + \chi)^{2/3} d^{-1/3} k^{-2} \, .
\end{equation} 
The non-dimensional  prefactors which we have omitted in (\ref{Second}) and (\ref{Spectra}) can be expressed in such a way that they can be determined from the constant $ C $ and the probability density function of $ \Delta u $. Thereby, the expressions are invariant under superposition of two fields, just as the corresponding expressions for Burgers turbulence.  

To compare the magnitudes of $ \epsilon $ and $ \chi $, we can use (\ref{DU}) and (\ref{DT}) to estimate both of them in terms of $ z $, 
\begin{eqnarray}
\epsilon   & \sim  & \left  (\frac{4}{3} \nu + \nu_{b} \right ) \frac{c_0^2}{\gamma^2 \delta x  d}  \langle z^2 \rangle_{\!A}  \, ,  \\
\chi  & \sim  &  \kappa \frac{ c_0^2(\gamma -1) }{\gamma^2 \delta x d} \langle z^2 \rangle_{\!A} \, ,
\end{eqnarray}
where $ \nu_b = \mu_b/\rho_0 $, $ \mu_b $ being the bulk viscosity.
Defining a Prandtl number as $ Pr \equiv (4\nu /3 + \nu_b)/ \kappa $, we see that  $ \chi \sim \epsilon $ if $ Pr \sim 1 $.  If this is the case, the  shock strength can be estimated as $ z \sim (\epsilon d)^{1/3}/c_0 $, the Mach number, $ M = \sqrt{1+ (\gamma+1)z/2\gamma} $, as $ M \sim 1+ (\epsilon d)^{1/3}/c_0 $ and the  shock width as in (\ref{Width}). The life time of a shock can be estimated as $ \tau \sim \delta x^{2}/\nu \sim \nu/(\epsilon d)^{2/3} $ and the ratio between the partial time derivate and the advective term in the entropy equation can thus be estimated as
 $ \partial_t S / c_0 \partial_x S \sim \delta x/ \tau c_0 \sim (\epsilon d)^{1/3}/c_0  $, motivating the assumption of quasi stationarity.

\section{Conclusions}
We showed that similar scaling relations hold for an acoustic field that is dissipated by weak shocks as for forced Burgers turbulence, with the difference that $ \epsilon $ should be replaced by $ \epsilon + \chi $, where $ \chi $ is associated with entropy production due to heat conduction.  From a principal point of view, the replacement of $ \epsilon $ by $ \epsilon + \chi $ makes a big difference. The third order structure function laws of incompressible turbulence  and Burgers turbulence are connected with the notion of a constant energy flux through scales, which is equal to $ \epsilon $.  It remains a theoretical challenge to investigate in what way
the quantity
$ \epsilon + \chi $ may be linked to an energy flux. From an experimental point of view the replacement of $ \epsilon $ by $ \epsilon + \chi $ is of less importance if $ Pr \sim 1 $. 
Let us assume that we set out to test the prediction of   \cite{KadomtsevPet1973} that an acoustic field is always dissipated by shocks if the Reynolds number is sufficiently large, and that we do this by generating a random acoustic field in a chamber with reflecting walls, 
using loud speakers whose total input power is $ P = 10^{-3} $ Watt/kg. Let us further assume that shocks are formed with a characteristic  distance $ d = 0.1 $ m. In a stationary state, we have $ \epsilon = P $ and given the relations derived in this paper we can estimate the relative pressure change over a shock as $ z \sim 10^{-4} $, the Mach number as $ M \sim 1 + 10^{-4} $, the chock velocity amplitude as $ \Delta u \sim  5 \, {\mbox{cm}}/ {\mbox{s}} $, the shock width as $ \delta x \sim 0.5 \,  {\mbox{mm}} $ and the Reynolds number as $ Re = \Delta u d/\nu \sim d/\delta x \sim  200 \gg 1$. Indeed, it would be an experimental challenge to produce an image of a field of such weak shocks, but it doesn't seem to be insurmountable. Likewise, the art of direct numerical simulations has developed into a stage in which it would be feasible to test the prediction of 
\cite{KadomtsevPet1973} by making a full Navier-Stokes simulation of a randomly forced acoustic field. It is the hope of the author that the present note will stimulate research along these lines.  
 \vskip 1cm
 
 \noindent The author would like to thank Gregory Falkovich for commenting on an early version of the manuscript.

\appendix

\section{Invariants of structure functions}
In this Appendix we derive relation (\ref{Isotropic}) and show that there are two invariants of the third order tensor structure function of an isotropic irrotational velocity field. The isotropic second order structure function can be written as
\begin{equation}
\langle \delta u_i \delta u_j \rangle = e_{i} e_{j} \langle \delta u_L^2  \rangle + s_{ij}  \langle \delta u_T ^2 \rangle \, ,
\end{equation} 
where $ {\bf e} = {\bf r}/r $ and $ s_{ij} = \delta_{ij} - e_i e_j $. By isotropy  $ \langle \delta u_L ^2  \rangle $ and $ \langle \delta u_T ^2 \rangle $ are functions of $ r $. By applying the irrotational  condition $ \epsilon_{ijk} \partial_j \langle \delta u_k \delta u_l \rangle = 0 $ and using $ \partial_{i} n_{j} = s_{ij}/r $ we obtain
\begin{equation}
\frac{1}{r} \epsilon_{ilk} e_k \left  ( \langle \delta u_L ^2 \rangle - \frac{{\mbox{d}}}{{\mbox{d}} r} (r  \langle \delta u_T ^2  \rangle ) \right ) = 0 \, ,
\end{equation} 
from which  (\ref{Isotropic}) follows. The third order tensor structure function can be written as
\begin{equation} \label{ThirdA} 
\langle \delta u_i  \delta u_j \delta u_k \rangle = 2(D_{ijk} + D_{jki} + D_{kij} ) \, ,
\end{equation} 
where $ D_{ijk} = \langle u_i u_j u_k^\prime \rangle $. Unprimed and primed quantities indicate positions  $ {\bf x} $ and  $ {\bf x} + {\bf r} $, respectively. In (\ref{ThirdA}) we have used that $ \langle u_i^\prime u_j^\prime u_k \rangle = - \langle u_i u_j u_k^{\prime} \rangle $ by isotropy and  $ \langle u_i^\prime u_j^\prime u_k^\prime \rangle =   \langle u_i u_j u_k \rangle $ by homogeneity.  Since the velocity field is irrotational we can write $ D_{ijk} = \langle \partial_i \phi \partial_j \phi \partial_k^\prime \phi^{\prime} \rangle $, where $ \phi $ is the velocity potential.  By homogeneity we have  
\begin{equation}  \label{F2} 
D_{ijk} = \frac{\partial B_{ij}} {\partial r_k} \, ,
\end{equation} 
where $ B_{ij} =  \langle \partial_i \phi \partial_j \phi  \phi^{\prime} \rangle $. The irrotational condition $ \epsilon_{ijk} \partial_j D_{lmk} = 0 $ is clearly fulfilled by (\ref{F2}).  By isotropy we can write $ B_{ij} = e_i e_j a(r) + s_{ij} b(r) $, where $ a(r) $ and $ b(r)$ are two scalar functions. 
It is quite clear that $ a(r) $ and $ b(r) $ are generally independent of each other since there is no further constraint on $ B_{ij} $.  A little bit of algebra gives
\begin{eqnarray}
\langle \delta u_L \delta u_L \delta u_L \rangle  & =  & 6 \frac{{\mbox{d}} a}{{\mbox{d}} r} \, , \\
\langle \delta u_L \delta u_T \delta u_T \rangle  & =  & 2 r^2 \frac{{\mbox{d}} }{{\mbox{d}}  r} \left ( \frac{b}{r^2} \right )  + \frac{4a}{r} \, .
\end{eqnarray} 
We conclude that the third order tensor structure function of an isotropic irrotational field  has two invariants.

\bibliographystyle{jfm}
\bibliography{acoustic}
\end{document}